\documentclass[journal,10pt]{IEEEtran}
\ifCLASSINFOpdf
  \usepackage[pdftex]{graphicx}
  \usepackage{amsthm,amsmath,amssymb}
\usepackage{makecell}
\usepackage{mathrsfs}
  \usepackage{float}
  \usepackage{algorithm}
\usepackage{algpseudocode}
\usepackage{amsmath}
\usepackage{multirow}
\usepackage{bm}
\usepackage{color}
\usepackage{comment}
 
\else
\fi
\usepackage{leftidx}
\usepackage{epstopdf}
\usepackage{amsmath}
\usepackage{subfigure}
\usepackage{makecell}
\usepackage{multirow}
\begin{document}

\title{LLM4MG: Adapting Large Language Model for Multipath Generation via Synesthesia of Machines}


 \author{Ziwei~Huang, Shiliang Lu, Lu~Bai, Xuesong~Cai,~and Xiang~Cheng
 
\thanks{The authors would like to thank Junlong Chen and Ruide Zhang for their help in the collection of multi-modal sensing-communication data.}
\thanks{Z.~Huang, X~Cai, and X.~Cheng are with the State Key Laboratory of Photonics and
Communications, School of Electronics, Peking University, Beijing, 100871, P. R. China (email: ziweihuang@pku.edu.cn, xuesong.cai@pku.edu.cn, xiangcheng@pku.edu.cn).}
\thanks{S. Lu is with the Joint SDU-NTU Centre for Artificial Intelligence Research (C-FAIR), Shandong University, Jinan, 250101, P. R. China, and also with the School of Software, Shandong University, Jinan 250101, P. R. China (e-mail: shiliang.lu@mail.sdu.edu.cn).}
\thanks{L. Bai is with the Joint SDU-NTU Centre for Artificial Intelligence Research (C-FAIR), Shandong University, Jinan, 250101, P. R. China, and also with the Shandong Research Institute of Industrial Technology, Jinan, 250100, P. R. China (e-mail: lubai@sdu.edu.cn).}}

\markboth{}
{Zeng \MakeLowercase{\textit{et al.}}: Bare Demo of IEEEtran.cls for IEEE Journals}

\maketitle

		\maketitle

\begin{abstract}
Based on Synesthesia of Machines (SoM), a large language model (LLM) is adapted for multipath generation (LLM4MG) for the first time. Considering a typical sixth-generation (6G) vehicle-to-infrastructure (V2I) scenario, a new multi-modal sensing-communication dataset is constructed, named SynthSoM-V2I, including channel multipath information, millimeter wave (mmWave) radar sensory data, RGB-D images, and light detection and ranging (LiDAR) point clouds. Based on the SynthSoM-V2I dataset, the proposed LLM4MG  leverages Large Language Model Meta AI (LLaMA) 3.2 for multipath generation via multi-modal sensory data. The proposed LLM4MG aligns the multi-modal feature space with the LLaMA semantic space through feature extraction and fusion networks. To further achieve general knowledge transfer from the pre-trained LLaMA for multipath generation via multi-modal sensory data, the low-rank adaptation (LoRA) parameter-efficient fine-tuning and propagation-aware prompt engineering are exploited. Simulation results demonstrate that the proposed LLM4MG outperforms conventional deep learning-based methods in terms of line-of-sight (LoS)/non-LoS (NLoS) classification with accuracy of $92.76\%$, multipath power/delay generation precision with normalized mean square error (NMSE) of $0.099$/$0.032$, and cross-vehicular traffic density (VTD), cross-band, and cross-scenario generalization. The utility of the proposed LLM4MG is validated by real-world generalization. The necessity of high-precision multipath generation for system design is also demonstrated by channel capacity comparison.
\end{abstract}

\IEEEpeerreviewmaketitle

\section{Introduction}
\IEEEPARstart {A}{n} in-depth understanding and precise modeling of the channel have been the cornerstone of the design and performance evaluation of communication systems throughout the generations \cite{cai1}. As a typical channel characteristic, channel small-scale fading, i.e., multipath fading, has a huge impact on instantaneous signal-to-noise ratio (SNR), bit error rate (BER), and channel capacity. Therefore, channel small-scale/multipath fading is a major factor influencing the transceiver design in communication systems and holds significant research value.

For conventional communication systems, i.e., from first-generation (1G) to fifth-generation (5G), the investigation of channel multipath fading provides support for physical layer transmission scheme design and offers a unified simulation validation platform for algorithms. In general, there are two conventional approaches to model  multipath fading characteristics, i.e., deterministic and stochastic channel modeling. Deterministic channel modeling, e.g., ray-tracing (RT) \cite{RTTT}, mimics the detailed procedure of physical radio propagation in a specific environment. Stochastic channel modeling, which is the recommended approach for standardized channel models \cite{3GPP,5GCM}, determines the channel parameter in a stochastic and low-complexity manner. Deterministic and stochastic modeling approaches have their advantages and disadvantages, and both can meet the requirement for the design of conventional communication systems, i.e., from 1G to 5G. However, future sixth-generation (6G) systems will deeply integrate  artificial intelligence (AI) into the core architecture design, forming an AI-native 6G system. Since the performance of AI-native  6G systems is fundamentally constrained by dataset scale and quality, the construction of a large-scale and high-quality channel dataset is of paramount importance. On one hand,  deterministic channel modeling suffers from significant  complexity with flawed  assumptions, rendering it inadequate for generating the large-scale channel dataset required. On the other hand, stochastic channel modeling is limited to generating low-precision data with key channel characteristics, failing to meet the high-quality demand. Therefore, the aforementioned deterministic and stochastic modeling approaches cannot generate large-scale and high-quality channel dataset and support  AI-native 6G system design. 

To address the aforementioned limitation, the powerful generative capabilities of AI models can be leveraged for efficient channel data generation. Note that the utilization of AI models naturally adapts to the AI-native 6G system design. A straightforward approach involves utilizing high-precision channel data as the basis for channel data generation via generative AI models. Based on real-world channel data, the authors in \cite{RF11} utilized the generative adversarial network (GAN) to generate channel data  under sub-6 GHz bands in static scenarios. To further consider millimeter wave (mmWave) bands in dynamic scenarios, the authors in \cite{RF22} utilized the generative model to generate path loss via RT-based channel data. Nonetheless, the aforementioned methods \cite{RF11,RF22} with uni-modal radio-frequency (RF) communications offers a limited understanding and characterization of propagation environment, and thus the  complexity is high and the precision is also limited \cite{myCOMST}.
Fortunately, in future 6G systems, communication modules and sensors will be equipped to obtain multi-modal information, such as RF channel, RF sensing, i.e., mmWave radar information, and non-RF sensing, i.e., RGB-D images and light detection and ranging (LiDAR) information \cite{AlkhateebA}--\cite{synthsom}. To adequately utilize the multi-modal information, inspired by  synesthesia of human, we proposed a novel framework of Synesthesia of Machines (SoM) \cite{SOM}. Based on integrated sensing and communications (ISAC) \cite{liufan1,liufan2} focused on RF communications and RF sensing, SoM \cite{SOM} considers the AI-native intelligent integration of RF communications and multi-modal sensing, including RF and non-RF sensing. 
With the help of the proposed SoM framework, more accessible sensory data with an in-depth understanding of environment and a comprehensive knowledge graph can be leveraged to efficiently achieve   cross-modal generation of channel data. As the theoretical foundation for cross-modal generation of channel data, the mapping mechanism between communications and multi-modal sensing  requires an extensive investigation~\cite{myCOMST}.
 
The mapping mechanism between communications and sensing for channel data generation has been currently investigated. By converting classification tasks into regression tasks, VGG-16 was leveraged to explore the sensing-communication mapping mechanism for path loss distribution generation from satellite images \cite{AhmadienO}.
To generate more detailed path loss data, the authors in \cite{Levie} proposed RadioUNet based on UNet structure and explored the mapping mechanism between city maps and path loss values. With reference to the input pre-processing way in RadioUNet \cite{Levie}, a PEFNet was proposed in \cite{JiangF} to explore the mapping mechanism and achieve path loss generation from city maps. Nevertheless, the aforementioned work in \cite{AhmadienO}--\cite{JiangF} failed to address the augmentation of raw sensory data, and thus the generation accuracy was limited. Through augmentation of environmental features in satellite images, based on convolutional neural networks (CNNs), the authors in \cite{QiuZ} explored the mapping mechanism between satellite images and communications for path loss generation. To acquire more precise environmental features from satellite images, the residual structure and attention mechanism were utilized, where path loss maps were generated via satellite images \cite{WangC}. Furthermore, the authors in \cite{GuptaA} extended the aforementioned work in \cite{AhmadienO}--\cite{WangC} focused on sub-6 GHz bands to mmWave bands, i.e., $28$~GHz, and generated path loss from LiDAR point clouds via CNNs. To further consider dynamic vehicular scenarios, based on our constructed M$^3$SC dataset \cite{dataset_cc}, the authors in \cite{WeiZ} explored the multi-modal sensing-communication mapping mechanism, which facilitates generating path loss distribution from RGB-D images and LiDAR point clouds.  However, the work in \cite{WeiZ} remains limited to generating  coarse-grained and large-scale channel data, failing to generate  fine-grained and small-scale channel data. 
Research on small-scale fading proves more challenging than large-scale fading \cite{BuiN}. In \cite{myWCL}, our previous work  preliminarily explored the mapping mechanism between LiDAR point clouds and small-scale fading, and achieved multipath scatterer generation from LiDAR point clouds in  vehicular scenarios. However, the work in \cite{myWCL} focused on the conventional deep learning model, i.e., multilayer perceptron (MLP), which exhibits two general limitations. On one hand, constrained by the limited inference capability of conventional deep learning models, the explored mapping mechanism fails to delve into fine-grained and small-scale fading, i.e., multipath fading. On the other hand, the conventional deep learning model requires retraining when adapting to new scenarios and frequency bands, thus significantly compromising deployment agility.

Compared to the conventional deep learning models, large language models (LLMs) possess more robust generation and generalization abilities \cite{LLM1}. The advent of LLMs has not only brought about a paradigm shift in natural language processing (NLP), but has also enhanced capabilities across multiple scientific and engineering disciplines, such as chemistry, biology, mathematics, and software engineering \cite{LLM65}. Currently, some work has exploited the in-context learning capacity of LLMs and has fine-tuned them for better domain adaptation to implement non-linguistic channel-related tasks in the physical layer, including channel perdition \cite{LLM4CP},  channel state information feedback \cite{LLM68}, and channel estimation \cite{LLM69}. However, the aforementioned work  \cite{LLM65}--\cite{LLM69} focused on \emph{uni-modal} RF channel information, which exhibits fundamental incompatibility with \emph{multi-modal} sensing-communication mapping mechanism exploration for cross-modal multipath generation. Consequently, there is an urgent need to leverage LLMs to explore the mapping mechanism for efficient multipath generation from multi-modal sensing under various scenarios and frequency bands. Nevertheless, adapting LLMs for multipath generation from multi-modal sensing poses substantial challenges. First, the substantial difference in data representation, acquisition frequency bands, and application objectives of channel multipath and multi-modal sensing  results in  complex and nonlinear mapping mechanisms. Second, the inherent divergence between linguistic representations in LLMs and multi-modal feature domains hinders knowledge transfer. Finally, the LLM-based method also imposes stringent requirements on the scale and quality of training datasets. 

To fill this gap, we consider a  6G vehicle-to-infrastructure (V2I) scenario and construct a new multi-modal sensing-communication dataset, named SynthSoM-V2I. Based on the SynthSoM-V2I dataset, we propose a novel LLM4MG method, which  for the first time adapts LLM for  multipath generation via SoM. The main contributions and novelties of this paper are summarized below.
\begin{enumerate}
\item LLM4MG is proposed as a novel method that adapts the LLM, i.e., Large Language Model Meta AI (LLaMA) 3.2, for cross-modal generation of fine-grained channel multipath parameters from multi-modal sensing via SoM. In the proposed LLM4MG, the complex and nonlinear mapping mechanism between multi-modal sensing and channel multipath is explored based on a new constructed SynthSoM-V2I dataset for the first time.
\item By achieving in-depth integration and precise alignment of AirSim \cite{AirSim}, WaveFarer \cite{WF}, and Sionna RT \cite{SRT}, we construct a new multi-modal sensing-communication V2I dataset, named SynthSoM-V2I. The SynthSoM-V2I dataset contains $211,395$ snapshots of RGB-D images, LiDAR point clouds, mmWave radar point clouds, and channel multipath data under high/low vehicular traffic densities (VTDs), mmWave/sub-6 GHz bands, and urban/suburban scenarios. Unlike datasets tailed for conventional deep learning models with a specific scenario/condition, the SynthSoM-V2I dataset meets the data requirement for LLM development and generalization evaluation
with various scenarios/conditions.
\item In the proposed LLM4MG, we employ multi-modal sensory feature extraction and fusion networks to obtain multi-modal features of the propagation environment  around transceivers. Furthermore, the multi-modal feature space is aligned with the LLaMA 3.2 semantic space. To efficiently achieve general knowledge transfer from the pre-trained LLaMA 3.2 to the mapping mechanism exploration for cross-modal generation of multipath data, the low-rank adaptation (LoRA) parameter-efficient fine-tuning and propagation-aware prompt engineering are utilized.
\item The proposed LLM4MG achieves superior performance compared to conventional deep learning models, attaining  line-of-sight (LoS)/non-LoS (NLoS) classification accuracy of $92.76\%$ while maintaining multipath power and delay generation normalized mean square error (NMSE) values of $0.099$ and $0.032$, respectively. A close fit between RT-based results and results based on the proposed LLM4MG according to channel statistical properties. The proposed LLM4MG also exhibits the most robust generalization capability across varying VTDs, frequency bands, and scenarios. The utility of the proposed LLM4MG is verified through real-world generalization. The necessity of high-precision multipath generation for system design is further shown by channel capacity comparison.
\end{enumerate}

The remainder of this paper is organized as follows. Section~II describes the constructed SynthSoM-V2I dataset. Section~III proposes the LLM4MG method to explore the multi-modal sensing-communication mapping mechanism for cross-modal multipath generation. Section~IV derives and analyzes the typical channel statistical properties.
Section~V presents simulation results, where the utility and necessity of high-precision multipath generation via the proposed LLM4MG are validated. Finally, Section~VI draws the conclusion.

\section{SynthSoM-V2I: A Multi-Modal Sensing-Communication Vehicle-to-Infrastructure Dataset}

In this section, we construct a new multi-modal sensing-communication V2I dataset, named SynthSoM-V2I. The SynthSoM-V2I dataset provides a reliable data foundation of the proposed LLM4MG. 
\begin{figure}[!t]
		\centering	
        \includegraphics[width=0.49\textwidth]{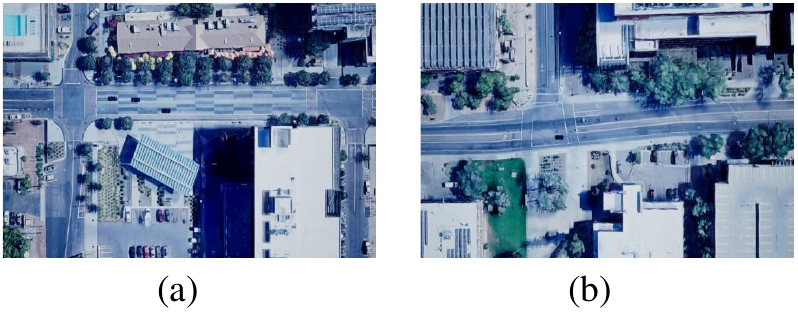}
    \caption{Urban and suburban scenarios in the SynthSoM-V2I dataset. (a) Urban scenario. (b) Suburban scenario.}
	\label{scenario}
	\end{figure}
    
\begin{figure*}[!t]
		\centering	\includegraphics[width=0.99\textwidth]{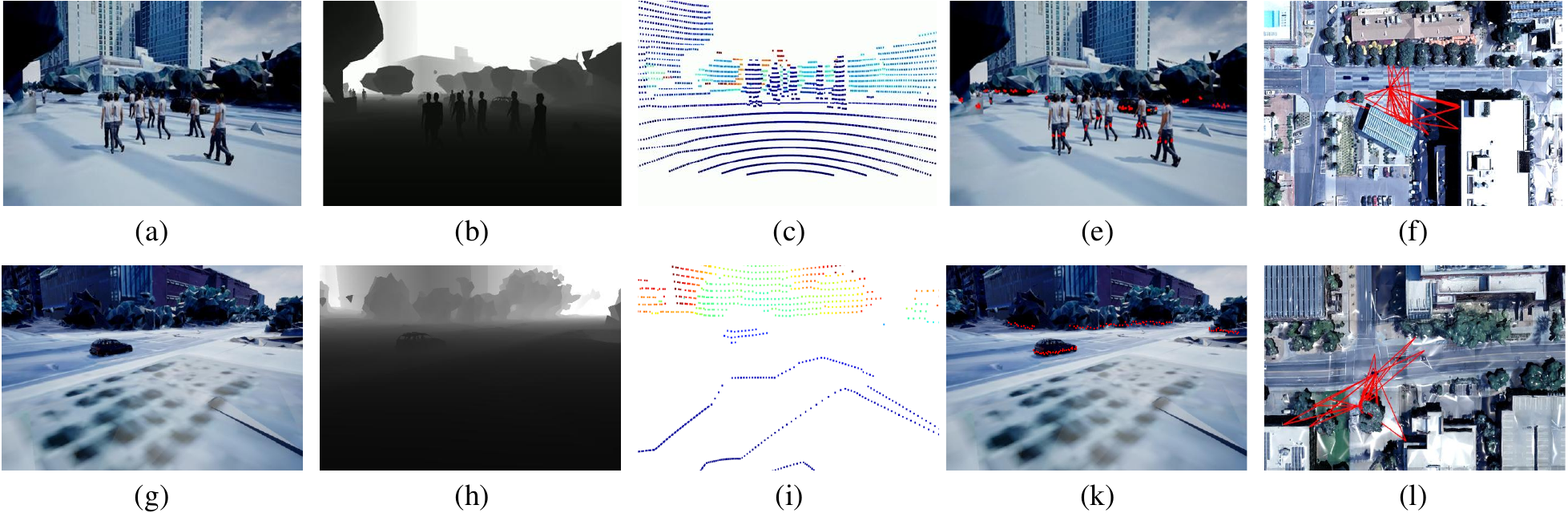}
	\caption{he constructed SynthSoM-V2I dataset using the urban scenario with low VTD at mmWave and the suburban scenario with low VTD as examples. Figs.~(a)--(e) are RGB image collected in AirSim, depth image collected in AirSim, denoised LiDAR point cloud collected in AirSim, mmWave radar point cloud collected in WaveFarer and registered to RGB image, and channel multipath data collected in Sionna RT under urban scenario with low VTD at mmWave, respectively. Figs.~(f)--(j) are RGB image collected in AirSim, depth image collected in AirSim, denoised LiDAR point cloud collected in AirSim, mmWave radar point cloud collected in WaveFarer  and registered to RGB image, and channel multipath data collected in Sionna RT under suburban scenario with low VTD at mmWave, respectively.}
	\label{dataset_1}
	\end{figure*}
    
\subsection{Dataset Description}
Building on the architecture and design principle of the SynthSoM dataset \cite{synthsom}, the SynthSoM-V2I dataset is constructed by utilizing three software, i.e., AirSim \cite{AirSim}, WaveFarer \cite{WF}, and Sionna RT \cite{SRT}, and achieves in-depth integration and precise alignment among them. Furthermore, based on the digital-twin technology utilized in the construction of our SynthSoM-Twin dataset \cite{junlong}, the SynthSoM-V2I dataset also creates a digital replica that is spatio-temporally consistent with the DeepSense 6G real-world scenario \cite{AlkhateebA}, and further collects multi-modal data, including RGB-D images, LiDAR point clouds, mmWave radar point clouds,
and channel multipath data.
To enhance the diversity of the SynthSoM-V2I dataset, we expand our previously constructed SynthSoM-Twin dataset \cite{junlong} by further incorporating cases with different VTDs and  frequency bands. 
Specifically, first, due to the huge impact of VTDs on vehicular channel characteristics based on the measurement and analysis \cite{VTD22}, we consider high and low VTDs in urban scenarios. 
Second, owing to the pronounced disparities in channel characteristics across frequency bands \cite{cai2}, we mimic two typical frequency bands, including sub-6 GHz and mmWave. In sub-6 GHz bands, the carrier frequency is $5.9$ GHz with $20$ MHz  bandwidth. In mmWave bands, the carrier frequency is $60$ GHz with $2$ GHz  bandwidth. Third, we consider two typical types of vehicular scenarios, i.e., urban and suburban, as shown in Fig.~\ref{scenario}.
Overall,  four different cases are considered in the SynthSoM-V2I dataset, including the urban scenario with low VTD at sub-6 GHz, urban scenario with high VTD at mmWave, urban scenario with low VTD at mmWave, and suburban scenario with low VTD at mmWave.

\subsection{Data Collection}
To collect multi-modal data, the vehicle and base station (BS) are equipped with multi-modal sensors and communication equipment.  For non-RF sensory data, the vehicle and BS collect RGB-D images and  LiDAR point clouds in AirSim. The collected RGB-D image is with $960 \times 540$ resolution. The LiDAR point cloud is collected by the LiDAR device that features $16$ channels with a scanning frequency of $20$ Hz, and is denoised by filtering. For RF sensory data, the vehicle and BS collect mmWave radar point clouds in WaveFarer. The mmWave radar operates in the frequency range of $77$~GHz to $78$~GHz, where the chirp length is $60~\mu$s and the sampling interval of echo signal is $0.2~\mu$s. For RF communication data, the vehicle is the transmitter (Tx) equipped with one antenna element of the phased array to realize omnidirectional transmission and the BS is the receiver (Rx) equipped with one antenna element, thus forming a single-input single-output (SISO) link. The link between the vehicle and BS contains channel multipath parameters, including LoS and NLoS paths with power and delay information in Sionna RT. 
The time interval between two snapshots is $33.33$ ms.
Overall, the SynthSoM-V2I dataset comprises $211,395$ snapshots of multi-modal sensing-communication data, including $42,279$ snapshots each of RGB-D images, LiDAR point clouds, mmWave radar point clouds, and channel multipath parameters for the development of the proposed LLM4MG.
For clarity, taking the urban scenario with low VTD at mmWave and the suburban scenario with low VTD at mmWave as examples, Fig.~\ref{dataset_1} depicts RGB-D images (RGB images and depth images), denoised LiDAR point clouds, mmWave radar point clouds, and channel information.

\section{LLM4MG: Adapting LLM for Multipath Generation in Sensing-Communication Channels}
\begin{figure*}[!t]
		\centering	\includegraphics[width=0.99\textwidth]{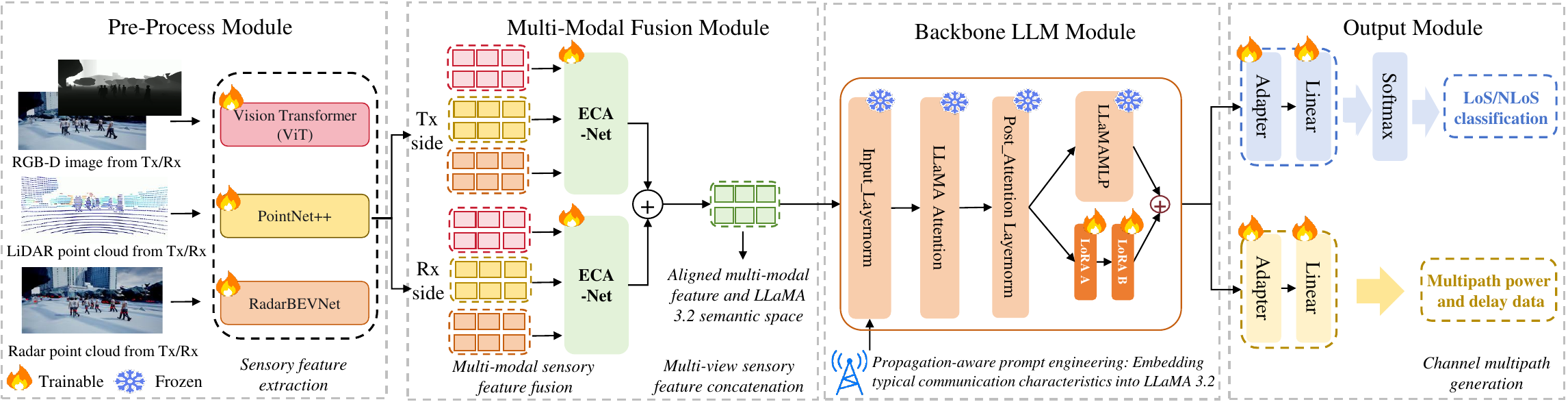}
	\caption{The proposed LLM4MG contains four main modules: (i) pre-process module; (ii) multi-modal fusion module; (iii) backbone LLM module with LoRA parameter-efficient fine-tuning and propagation-aware prompt engineering; (iv) output module.}
	\label{framework}
	\end{figure*}

Based on the SynthSoM-V2I dataset, a novel method, named LLM4MG, is proposed, which leverages the LLM, i.e., LLaMA 3.2, for multipath generation in sensing-communication channels, incorporating LoRA-based parameter-efficient fine-tuning and propagation-aware prompt engineering. The  framework of the proposed LLM4MG is shown in Fig.~\ref{framework}. The proposed LLM4MG explores the complex and nonlinear mapping mechanism between multi-modal sensing and channel multipath fading, thus achieving the cross-modal generation of fine-grained channel multipath parameters for the first time. A detailed explanation of the module and the training process for the proposed LLM4MP are given below.

\subsection{Pre-Process Module}
To extract the sensory feature, a pre-process module is introduced to process the input data. The input data contains multi-modal sensory information, such as RGB-D images, LiDAR point clouds, and mmWave radar point clouds, collected from multiple views, such as Tx (vehicle)  and Rx  (BS).  By employing canonical sensory   feature extraction networks, the environmental feature around transceivers can be efficiently extracted by the designed pre-process module.

First, Vision Transformer (ViT) in \cite{vit} is utilized to extract the feature of non-RF sensory RGB-D images. By splitting an image into non-overlapping patches and treating them as a token sequence, ViT leverages a Transformer encoder with self-attention to model global relationships between all patches simultaneously. Second, to extract the feature of non-RF sensory LiDAR point clouds, we exploit PointNet++, which is a typical neural network architecture for 3D point cloud processing \cite{PointNet}. The architecture of PointNet++ utilizes a set abstraction mechanism that systematically downsamples LiDAR point clouds while preserving local geometric patterns through multi-scale neighborhood grouping and hierarchical PointNet feature extraction. Third, we leverage an advanced network, i.e., RadarBEVNet in RCBEVDet \cite{RadarBEVNet}, to extract the feature of mmWave radar point clouds. RadarBEVNet integrates a dual-stream radar backbone network with a radar cross-section (RCS) aware bird's eye view (BEV) encoder. RadarBEVNet leverages a hybrid point-based and transformer-based encoder to process sparse mmWave radar point clouds, where cross-attention mechanisms dynamically update mmWave radar point features while incorporating RCS characteristics. Overall, the multi-modal sensory data collected from the vehicle and BS are properly processed through the aforementioned feature extraction networks.

\subsection{Multi-Modal Fusion Module}
To obtain the comprehensive environmental information around the transceiver, the extracted multi-modal sensory feature from Tx and Rx needs to be properly fused. Towards this objective, the multi-modal fusion module is designed, which consists of multi-modal sensory
 feature fusion and multi-view
feature concatenation. 

For the multi-modal sensory feature fusion, the multi-modal sensory feature from Tx/Rx is fused by a low computational complexity network, i.e., efficient channel attention network (ECA-Net) \cite{ECA-Net}. The conventional channel attention mechanism, e.g., squeeze-and-excitation network (SENet) \cite{SENet}, captures interactions among all channels through fully connected layers, resulting in highly computational complexity. In contrast, the ECA-Net employs one-dimensional (1D) convolution with a kernel size of $k$ to achieve local cross-channel interaction, which solely considers relationships between adjacent $k$ channels. The kernel size $k$ is defined as 
\begin{equation}
    k=\psi(C)=\left|\frac{\log _2(C)}{\gamma}+\frac{b}{\gamma}\right|_{\text {odd }}
\end{equation}
where $C$ is the channel dimension and  $|t|_\text{odd}$ is the nearest odd number of $t$. $\gamma$ and $b$ are parameters of mapping function, which can be typically set to $\gamma = 2$ and $b=1$. The parameter quantity and computational overhead can be significantly reduced via the ECA-Net. Before utilizing ECA-Net, the features of RGB-D images, mmWave radar point clouds, and LiDAR point clouds are concatenated along the channel dimension. Subsequently, this feature is fed into the ECA-Net module. The ECA-Net module adaptively learns the local dependencies between channels through lightweight 1D convolutions, thus performing importance weighting on the channels of different modalities to achieve effective fusion and enhancement of multi-modal sensory features.
For the multi-view feature concatenation, the processed Tx-view features and Rx-view features from ECA-Net undergo a concatenation operation, i.e., the multi-modal sensory features from the Tx and Rx sides are concatenated along the channel dimension. As a consequence,  the comprehensive environmental feature encompassing Tx and Rx sides can be obtained. 

In summary, based on the multi-modal sensory feature fusion via ECA-Net and multi-view sensory feature concatenation via concatenation operation, the feature dimension is 2048, which is aligned with the input embedding dimension of LLaMA 3.2 semantic space.

\subsection{Backbone Large Language Model Module}

The backbone LLM module is important to process the
representations extracted by the multi-modal fusion module. For the exploration of mapping mechanism, we utilize a Meta's state-of-the-art open-weight generative AI model, i.e., LLaMA 3.2 \cite{llama3}. The motivation of selecting LLaMA 3.2 is given below. On one hand, building upon its predecessors, LLaMA 3.2 features enhanced architecture with optimized transformer-based layers and improved tokenization efficiency. As a consequence, LLaMA 3.2 is a multi-modal large language model (MLLM) naturally suited for tasks of multi-modal sensing-communication mapping mechanism exploration for multipath generation.  On the other hand, LLaMA 3.2 is of robust reasoning capabilities, enabling its potential to explore complex and nonlinear mapping mechanism between multi-modal sensing and channel multipath parameters. To achieve general knowledge transfer from the pre-trained LLaMA 3.2 for multipath generation task by exploring the mapping mechanism, the LoRA parameter-efficient fine-tuning and propagation-aware prompt engineering are utilized.

For the LoRA parameter-efficient fine-tuning,  we utilize it to enhance the performance of the pre-trained LLaMA 3.2  on the task related to multipath generation by exploring the mapping mechanism. Specifically, we aim to train two low-rank matrices in the model’s feed-forward network. The pre-trained weight is $W_0 \in \mathbb{R}^{d_{\text {out}} \times d_{\text {in }}}$ with output dimension $d_{\text {out}}$ and input dimension $d_{\text {in}}$. Furthermore, two trainable low-rank matrices are $B \in \mathbb{R}^{d_{\text {out }} \times r}$ and $A \in \mathbb{R}^{r \times d_{\mathrm{in}}}$ with $r \ll \min \left(d_{\mathrm{out}}, d_{\mathrm{in}}\right)$. As a result, the fine-tune weight $W \in \mathbb{R}^{d_{\text {out}} \times d_{\text {in}}}$ is given as 
\begin{equation}
    W=W_0+\frac{\alpha}{r} B A
\end{equation}
where $r$ is the rank of the low-rank approximation and $\alpha$ is a hyperparameter, which supports the modification  of the rank $r$. Given the input to the feed-forward network $x_t$ and the output $y_t$, the forward propagation of the model is given by
\begin{equation}
y_t=W x_t=W_0 x_t+\frac{\alpha}{r} B A x_t.
\end{equation}

For the propagation-aware prompt engineering, typical communication characteristics, including carrier frequency, bandwidth, transceiver distance, azimuth and elevation angles of transceiver antennas, are embedded into LLaMA 3.2. Specifically, we input typical communication characteristics as text into LLaMA 3.2's tokenizer in the form of prompts to perform tokenization on the prompt text. For a text segment describing communication information, the tokenizer decomposes it into corresponding token sequences according to its internal rules and vocabulary. In LLaMA 3.2, each token corresponds to a unique embedding vector. When tokens are input into the model, the model retrieves the corresponding embedding vectors from the embedding layer based on the token indices, thus transforming the text prompts into feature vectors. These feature vectors serve as the foundation for the model's subsequent mapping mechanism exploration, which can represent the semantic information of the input text.  

Overall, LoRA is integrated into the linear layer within the feed forward network (FFN) of LLaMA 3.2, while keeping the remaining parameter frozen. As a result, the trainable parameter of LLaMA  3.2 is exceedingly reduced, thus lowering the training cost and enhancing training efficiency. The utilization of prompt engineering further embeds communication information into LLaMA 3.2, endowing the model with propagation comprehension capabilities.

\subsection{Output Module}
For downstream cross-modal generation tasks, the output is configured to classify the  LoS/NLoS status in channels, as well as to generate the multipath power and delay. The LoS/NLoS classification and multipath power and delay generation are of paramount importance. Specifically, LoS/NLoS classification facilitates blockage prediction and beamforming. For the multipath power and delay generation, it can support the successful design of communication systems and RF-based positioning by generating time-varying power delay profile (PDP) for spatially-consistent channels. Certainly, the channel evolves continuously and consistently with changes in the spatial positions of the Tx/Rx, which is referred to as spatial consistency \cite{SCC1}--\cite{SCC3}. The capturing of spatial consistency is a key requirement for channel modeling in 6G systems. For a propagation path, a spatially consistent channel has a relatively short stationary distance/time, within which its power and delay remain largely unchanged whereas its phase undergoes significant changes. This is in agreement with the real propagation geometrical characteristics, where limited bandwidth struggles to capture small delay variations, and power fading caused by small distance changes can be neglected. 
By contrast, the high carrier frequency means that movement on the order of a wavelength can induce significant phase variations. Consequently, the phase of each path can be regarded as randomly varying within the stationary distance/time. Owing to the extreme phase sensitivity, accurately recovering the instantaneous channel impulse response (CIR) $h(t,\tau)$ with exact phase from Tx/Rx positions and multi-modal information is significantly difficult. Consequently, we first generate a time-varying PDP within the stationary distance/time. Deriving the considered SISO CIR from the time-varying PDP is then straightforward, i.e., we assign a random phase, e.g., a uniform distribution over $[0,2\pi)$ \cite{Yang22}, to each path. The effectiveness of randomly generating multipath phase will be validated in Section V-D.

The standard LLM generally transforms the output features of Transformer block into a probability distribution over the vocabulary, and then select the token with the highest probability as the predicted output. Nevertheless, for cross-modal generation tasks, i.e., LoS/NLoS classification and multipath power/delay generation via multi-modal sensory data, the output is generally difficult to represent in text. Furthermore, the increase in the vocabulary size leads to the fact that this mapping incurs in a storage and computational cost. For example, LLaMA 3.2' vocabulary of 128K words needs an output layer with at least 128K dimensions. This does not match the output space for generating physical channel-related information composed of low-dimensional and extremely sparse discrete variables. As a result, retaining the original output of LLaMA 3.2 not only requires coarse quantization of continuous values in the vocabulary with unavoidable truncation errors, but also leads to significant redundant computational overhead.

To address the aforementioned challenge, we design a lightweight adapter tailored for cross-modal generation tasks, including LoS/NLoS classification task and multipath power/delay generation task. The designed adapter aims to adequately acquire the target output related to the task, and thus the  performance is improved and the resource demand related to large vocabulary sizes is reduced. Specifically, the adapter is connected directly to the output of LLaMA 3.2, thus aligning the task’s output feature vector with the semantic space of LLaMA 3.2. For the task $p$, i.e., LoS/NLoS classification task or  multipath power/delay generation task, assume that the  output feature of LLaMA 3.2  is $\boldsymbol{X}_p^\text{LLaMA}$ and the adapter is $\text{Adapter}_p^{\text{out}}$. The alignment between output feature vector and semantic space of LLaMA 3.2 for the task $p$ is expressed by 
\begin{equation}   \boldsymbol{X}_p^{\text{mapping}}=\text{Adapter}_p^{\text{out}}\left(\boldsymbol{X}_p^\text{LLaMA}\right)
\end{equation}
where $\boldsymbol{X}_p^{\text{mapping}}$ represents the output of the adapter for the task $p$. The design specifications of the adapter can be elaborated as follows. First, the temporal feature output by the LLaMA 3.2 decoder layer undergoes global average pooling to derive a fixed 2048-dimensional cross-modal representation. Subsequently, such a representation is fed in parallel into different residual branches due to  different types of tasks, i.e., classification and generation. One of the  residual branch constitutes a LoS/NLoS classifier, which employs two 512-dimensional ReLU fully-connected layers with residual connections to project features into two-dimensional (2D) logits. Other residual branches consist of residual branches tailored for the generation of multipath power and delay values, with each layer employing 512-dimensional hidden units and 0.3 Dropout for regularization. In addition, the feature output by the adapter is processed through the linear layer for the task $p$, which can be given as 
\begin{equation}   \boldsymbol{X}_p^{\text{output}}=\text{Linear}\left(\boldsymbol{X}_p^{\text{mapping}}\right)
\end{equation}
where $\boldsymbol{X}_p^{\text{output}}$ denotes the output result for the task $p$. Finally, for the LoS/NLoS
classification task, to convert the network output into a class-probability distribution, the Softmax function is utilized and is written as 
\begin{equation}
y_\text{classification}=\text{Softmax}\left(\boldsymbol{X}_p^{\text {output }}\right)
\end{equation}
where $y_\text{classification}$ denotes the probability distribution of LoS and NLoS.

\subsection{Training Configuration}
The proposed LLM4MG is developed via the SynthSoM-V2I dataset, where dataset is divided into the training set, validation
set, and testing set in the proportion of $3:1:1$. Furthermore, the proposed LLM4MG utilizes a three-stage training scheme. The first stage is the warm-up stage, where the first 3 epochs are utilized for linear warm-up, with the learning rate increasing linearly from $10\%$ of the initial learning rate to the full initial learning rate. The motivation underlying warm-up is to allow LLaMA 3.2 to start learning smoothly in the early stages of training, avoiding potential instability caused by excessively large parameter updates due to the high learning rate in the early stages of training. The second stage employs a cosine annealing scheduler to dynamically adjust the learning rate. During the first two stages, the LLaMA 3.2 parameter remains frozen, which means that these parameters are not trainable. In the third stage, LoRA is activated to fine-tune LLaMA 3.2 while keeping the other parameters trainable after $10$ epochs, which can achieve the decent mapping mechanism exploration performance via generalized representations. All three stages utilize the same loss function, which is given by 
\begin{equation}
    \text{Loss}=\sum_p f_{\text {loss},p}\left(\mathbf{X}_p^\text{output}, \mathbf{X}_p^\text{gr}\right)
\end{equation}
where $f_{\text {loss},p}$ is the loss function of the task $p$ and $\mathbf{X}_p^\text{gr}$ is the ground truth of the task $p$. The loss function $f_{\text {loss},p}$ is designed to consider the feature of each task and the propagation effect. For the LoS/NLoS classification task, cross-entropy loss function is exploited. For the multipath power/delay generation task, we utilize the NMSE as the loss function and further embed the propagation effect. The NMSE of the multipath power/delay generation  is written as 
\begin{equation}
    \text{NMSE}_\text{power}=\frac{\sum_{n=1}^N \mu_n\left(\omega_n-\hat{\omega}_n\right)^2}{\sum_{n=1}^N \omega_n^2}
\end{equation}
\begin{equation}
    \text{NMSE}_\text{delay}=\frac{\sum_{n=1}^N \left(\epsilon_n-\hat{\epsilon}_n\right)^2}{\sum_{n=1}^N \epsilon_n^2}
\end{equation}
where ${\omega}_n$/$\epsilon_n$ is the ground truth of the $n$-th propagation path power/delay parameter, $\hat{\omega}_n$/$\hat{\epsilon}_n$ is the  $n$-th generated propagation path power/delay parameter, and $\mu_n$ is the weight of power generation for the $n$-th propagation path. Based on the propagation effect, the most dominant propagation path with the strongest power accounts for a substantial proportion of the total received power. In such a condition, accurate generation of the most dominant propagation path is of paramount importance. As a consequence, the weight of power generation for the most dominant propagation path is set to $\mu=3$. The remaining weight is set to $\mu=1$. 

\section{Channel Statistical Properties}
In this section, we derive and analyze the key channel statistical properties, such as PDP, root mean square (RMS) delay spread, and frequency correlation function (FCF).

\subsection{Power Delay Profile}
The time-variant PDP $\Omega(t,\tau)$ characterizes a channel multipath propagation by quantifying the received signal power distribution across different delays. This fundamental metric reveals the power
and delay of multipath components and distinct peaks corresponding to dominant propagation paths, e.g., the $n$-th propagation path, with the delay $\tau_n(t)$ and complex channel gain $h_n(t)$ at the snapshot $t$, which can be given as 
\begin{equation}
    \Omega(t,\tau) = \sum_{n=1}^{N} \left |h_n(t) \right |^2 \delta(\tau - \tau_n(t)) 
\end{equation}
where $N$ is the number of propagation paths.  The variation of PDPs is caused by evolutions of multipath components, which can support RF-based positioning.

\subsection{Root Mean Square Delay Spread}
The RMS delay spread, defined as the square root of the second central moment of the time-variant PDP $\Omega(t,\tau)$, serves as a fundamental metric for quantifying channel delay dispersion in wireless communications. The RMS delay spread $\tau_\text{RMS}(t)$ can be written as 
\begin{equation}
    \tau_{\mathrm{RMS}}(t)=\sqrt{\frac{\sum_{n=1}^{N} \Omega\left(t, \tau_n\right) \tau^2_n(t)}{\sum_{n=1}^{N} \Omega\left(t, \tau_n\right)}-\bar{\tau}(t)^2}
\end{equation}
where $\tau_n(t)$ denotes the delay of the $n$-th propagation path. Furthermore, $\bar{\tau}(t)$ is the mean delay at the snapshot $t$ and can be given as 
\begin{equation}
 \bar{\tau}(t)   = \frac{\sum_{n=1}^{N} \Omega\left(t, \tau_n\right) \tau_n(t)}{\sum_{n=1}^{N} \Omega\left(t, \tau_n\right)}.
\end{equation}
The acquirement of RMS delay spread facilitates the  determination of system architecture and modulation schemes.

\subsection{Frequency Correlation Function}
The FCF $\xi(t;f,\Delta f)$  quantifies the statistical correlation between a channel's transfer function at two frequencies separated by $\Delta f$. The FCF is derived as the Fourier transform of the PDP $\Omega(t,\tau)$ in respect of $\tau$ and can be given as
\begin{equation}
    \xi(t;f,\Delta f)=\int_{-\infty}^{+\infty} \Omega(t,\tau) e^{-j 2 \pi \Delta f \tau} d \tau .
\end{equation}
The analysis of FCF can support the design of  frequency-domain equalizers, which can be utilized to mitigate the effects of multipath fading and improve signal reception.

\section{Simulation Results and Analysis}
In this section, the simulation setup is given and the performance of the proposed LLM4MG is evaluated from various perspectives, including overall performance, generalization ability, as well as efficiency/complexity evaluation. Furthermore, the ablation experiment is conducted to demonstrate the contribution of each module in the framework. Finally, the utility of the proposed LLM4MG is validated by real-world generalization evaluation via Real2Real, Sim2Real, and Mixed2Real testing and the necessity of high-precision multipath generation for system design is demonstrated by channel capacity comparison via Shannon formula. 

\subsection{Simulation Setup}

\subsubsection{Dataset Overview}
The SynthSoM-V2I dataset creates a digital replica that is spatio-temporally consistent with the real-world scenario in the DeepSense 6G dataset \cite{AlkhateebA} and collects $211,395$ snapshots, containing RGB-D images, LiDAR point clouds, mmWave radar point clouds, and channel multipath data.  To ensure the dataset diversity, four different cases are considered in the SynthSoM-V2I dataset, including the urban scenario with low VTD at sub-6 GHz, urban scenario with high VTD at mmWave, urban scenario with low VTD at mmWave, and suburban scenario with low VTD at mmWave.

\subsubsection{Baselines}
To demonstrate the superiority of the proposed
LLM4MG, three conventional deep learning models, including MLP, ResNet, and Transformer,  are regarded as baselines. In the simulation, the backbone LLM module is replaced with MLP, ResNet, and Transformer for comparison.
\begin{itemize}
  \item \emph{MLP}: MLP consists of multiple layers of interconnected nodes. Since MLP is capable of learning complex non-linear patterns through backpropagation and gradient descent, it is widely utilized in channel-related tasks \cite{MLP11}.
  \item  \emph{ResNet}: ResNet is a deep CNN architecture, which introduces residual block to address degradation. ResNet leverages batch normalization and shortcut connections to stabilize training, thus widely utilizing in channel-related tasks \cite{ResNet11}. Here, ResNet-34 is utilized for comparison. 
 \item  \emph{Transformer}: Transformer is a revolutionary deep learning architecture that relies on self-attention mechanisms and can be properly utilized in channel-related tasks, e.g., a transformer-based parallel channel predictor developed in \cite{Transformer11} with the capability of mitigating error propagation. 
\end{itemize}

\subsubsection{Network and Training Parameters}
According to the measurements \cite{path11,path22}, the number of dominant propagation paths in the environment is limited, e.g., 6 paths. In this case, we set the number of  generated propagation paths to $N=6$, where the multipath information accounts for 90\% of the power.   
In the ECA-Net, kernel size is set to $k=3$. Considering the trade-off between multipath generation accuracy and model complexity, the decoder number of LLaMA 3.2 is set to 2. For the setting of the LoRA fine-tuning method, we set $r = 8$ and $\alpha = 32$. Both the warm-up and cosine annealing scheduler are employed to train LLM4MG. The first 3 epochs serve as the warm-up phase, where the learning rate increases linearly from the minimum value of $1 \times 10^{-6}$ to $1 \times 10^{-5}$. In subsequent training phases, the learning rate undergoes dynamic adjustment via a cosine annealing scheduler. Other hyperparameters for
model training are listed in Table~\ref{hyperparameters}.

\begin{table}[!t]
		\centering
        \renewcommand\arraystretch{1.3}
		\caption{Hyperparameter Setting.}
		\begin{tabular}{|c|c|}
			\hline
		\makecell[c]{\textbf{Hyperparameter}}	 &
	\makecell[c]{\textbf{Setting}} 	  	\\
			\hline
		Batch size	& $24$ \\ 
	\hline
 		Epochs & $100$ \\ 
	\hline
  		Optimizer	& AdamW \\ 
	\hline
    Learning rate scheduler	& Cosine Annealing \\ 
    \hline
      Cosine annealing period	& 80 Epochs (Epochs $\times~0.8$) \\
      \hline
      Learning rate range	& $\left[5 \times 10^{-7}, 1 \times 10^{-5}\right]$\\ 
	\hline
		\end{tabular}	
		\label{hyperparameters}
	\end{table}

\subsubsection{Performance Metric}
For the LoS/NLoS classification task, we utilize the classification accuracy $\text{A}_\text{cls}$, which can be given as 
\begin{equation}
    \text{A}_\text{cls} = N_\text{accurate}/N_\text{all}
\end{equation}
where $N_\text{accurate}$ is the number of accurately classified samples and $N_\text{all}$ is the number of all samples. For the multipath power/delay generation task, normalized mean absolute error (NMAE) and NMSE are leveraged to measure the error between the generated multipath power/delay and the ground truth.  On one hand, the advantage of NMAE lies in its ability to provide a scale-invariant error measure and an intuitive representation of absolute generation errors. On the other hand, due to the extremely small values of the generated multipath power and delay, even minor errors can lead to significant differences. As a complement, NMSE, which penalizes larger errors more heavily, is further introduced for evaluation. Therefore, the simultaneous utilization of NMAE and NMSE not only provides provide a scale-invariant error measure but also effectively quantifies the impact of outliers. 

Let $\hat{P}_{\mathrm{G}}(t,n)$/$\hat{D}_{\mathrm{G}}(t,n)$ represent the generated multipath power/delay parameters and ${P}_{\mathrm{GT}}(t,n)$/${D}_{\mathrm{GT}}(t,n)$ represent the corresponding ground truth of the $n$-th propagation path at the snapshot $t$. The performance metrics NMAE and NMSE of the multipath power and delay generation can be given by 
\begin{equation}
\operatorname{NMAE}_\text{pt}=\sum_{t=1}^T\frac{ \sum_{n=1}^N |{P}_{\mathrm{G}}(t,n)-\hat{P}_{\mathrm{G}}(t,n)|}{T\sum_{n=1}^N {P}_{\mathrm{G}}(t,n)}
\end{equation}
\begin{equation}
\operatorname{NMAE}_\text{dt}=\sum_{t=1}^T\frac{ \sum_{n=1}^N |{D}_{\mathrm{G}}(t,n)-\hat{D}_{\mathrm{G}}(t,n)|}{T\sum_{n=1}^N {D}_{\mathrm{G}}(t,n)}
\end{equation}
\begin{equation}
\operatorname{NMSE}_\text{pt}=\sum_{t=1}^T\frac{ \sum_{n=1}^N |{P}_{\mathrm{G}}(t,n)-\hat{P}_{\mathrm{G}}(t,n)|^2}{T\sum_{n=1}^N {P}_{\mathrm{G}}(t,n)^2}
\end{equation}
\begin{equation}
\operatorname{NMSE}_\text{dt}=\sum_{t=1}^T\frac{ \sum_{n=1}^N |{D}_{\mathrm{G}}(t,n)-\hat{D}_{\mathrm{G}}(t,n)|^2}{T\sum_{n=1}^N {D}_{\mathrm{G}}(t,n)^2}
\end{equation}
where $\operatorname{NMAE}_\text{pt}$,
$\operatorname{NMAE}_\text{dt}$,
$\operatorname{NMSE}_\text{pt}$, and $\operatorname{NMSE}_\text{dt}$ represent the NMAE of multipath power generation, NMAE of multipath delay generation, NMSE of multipath power generation, and NMSE of multipath delay generation, respectively. $T$ denotes the number of total snapshots and  $N$ denotes the number of generated propagation paths, i.e., $N=6$.

\subsection{Performance Evaluation}

\begin{table*}[!t]
		\centering
        \renewcommand\arraystretch{1.3}
		\caption{Performance of the Proposed LLM4MG and Other Baselines Under the Urban Scenario with Low VTD at mmWave, Where \textbf{Boldface} Indicates The Best Result And \underline{Underlining} Denotes The Second-Best Result.}
		\begin{tabular}{|c|c|c|c|c|}
			\hline
		\makecell[c]{\textbf{Result}}	 &
	\makecell[c]{\textbf{LLM4MG}} &
	\makecell[c]{\textbf{MLP}} 	 &
	\makecell[c]{\textbf{ResNet}}   &
	\makecell[c]{\textbf{Transformer}}   	\\
			\hline
		LoS/NLoS classification accuracy	& $\textbf{92.76}\%$ & $91.89\%$ & $89.67\%$ & $\underline{91.94}\%$\\ 
        \hline
        NMAE of multipath power generation	& $\textbf{0.218}$ & $0.295$ & $\underline{0.265}$ & $0.317$\\ 
                \hline
        NMSE of multipath power generation	& $\textbf{0.081}$ & $0.130$ & $0.130$ & $\underline{0.129}$\\ 
                \hline
        NMAE of multipath delay generation	& $\textbf{0.099}$ & $0.240$ & $\underline{0.202}$ & $0.270$\\ 
                \hline
        NMSE of multipath delay generation	& $\textbf{0.032}$ & $\underline{0.090}$ & $0.106$ & $0.119$\\ 
	\hline
		\end{tabular}	
		\label{overall}
	\end{table*}

\subsubsection{Overall Performance}
For comparative evaluation, we conduct training utilizing the urban scenario with low VTD at mmWave in the SynthSoM-V2I dataset.
In Table~\ref{overall}, the proposed LLM4MG outperforms the conventional deep learning-based methods, including MLP, ResNet, and Transformer, for the LoS/NLoS classification task and the multipath power/delay generation task. This is attributable to the utilization of general knowledge of pre-trained LLaMA and its interference ability, enhancing feature representation. However, conventional deep learning models exhibit limited performance in high-mobility vehicular scenarios due to their constrained interference abilities. Specifically, for the LoS/NLoS classification task, the proposed LLM4MG  achieves a classification accuracy of $92.76\%$, demonstrating the best performance.  For the multipath power generation task via the proposed LLM4MG, as listed in Table~\ref{overall}, the NMAE and NMSE of multipath power ratio accuracy are $0.218$ and $0.081$, respectively. Compared to the more pronounced multipath power, the generation of multipath delay exhibits lower NMAE and NMSE, i.e., $0.099$ and $0.032$ via the proposed LLM4MG, respectively. In comparison, the proposed LLM4MG demonstrates significant improvements over the baselines, which can achieve over $2.02$ dB enhancement in power generation accuracy and over $4.49$ dB enhancement in delay generation accuracy for the NMSE metric. Therefore, compared to the LoS/NLoS classification task, the proposed LLM4MG demonstrates more significant advantages over conventional deep learning models in the more complex task of multipath power and delay generation.

\begin{figure}[!t]
		\centering	\includegraphics[width=0.49\textwidth]{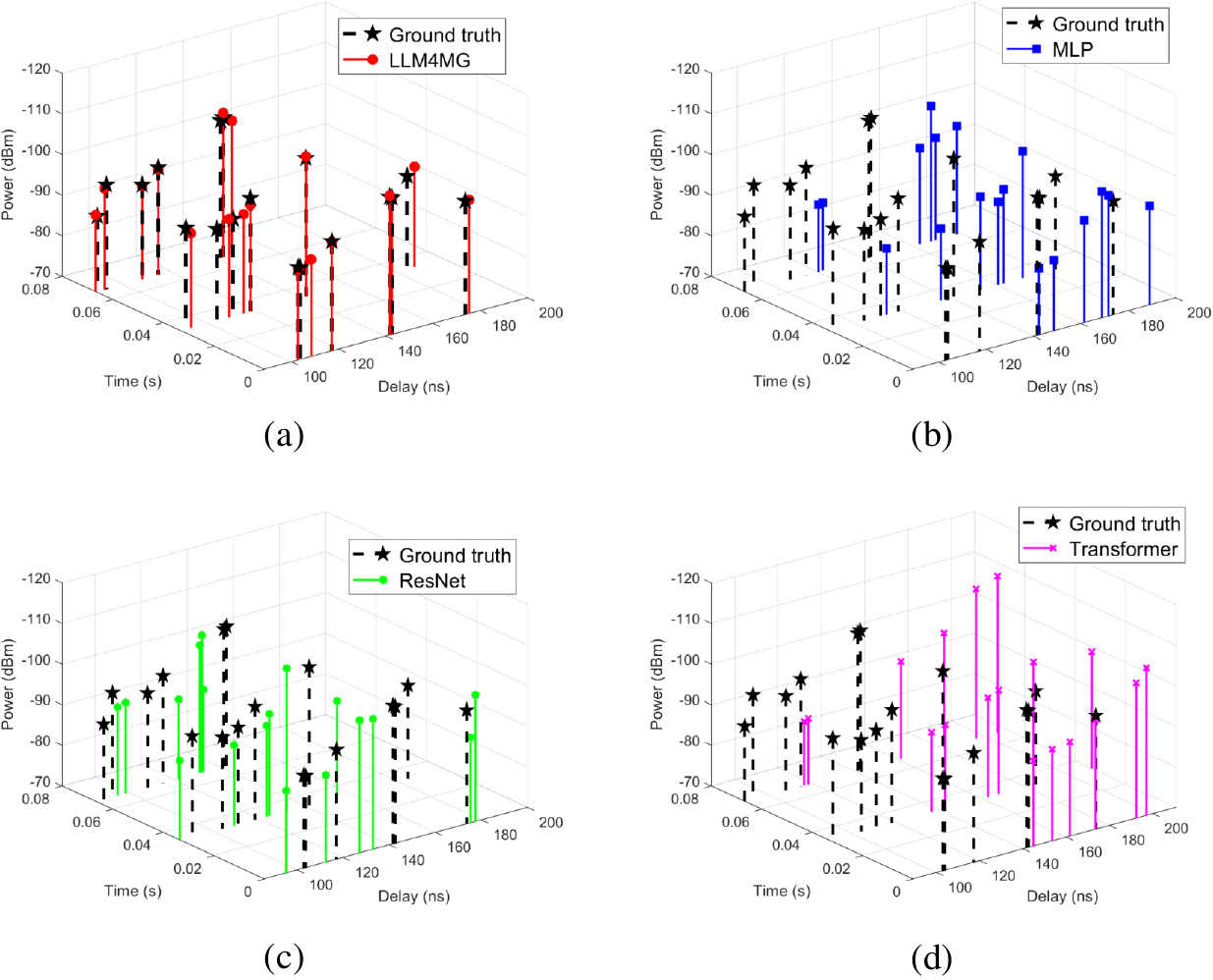}
	\caption{Time-varying PDPs for spatially-consistent channels using 3 snapshots as an example. (a) LLM4MG. (b) MLP. (c) ResNet. (d) Transformer.}
	\label{PDP}
	\end{figure}

\begin{figure}[!t]
		\centering	\includegraphics[width=0.45\textwidth]{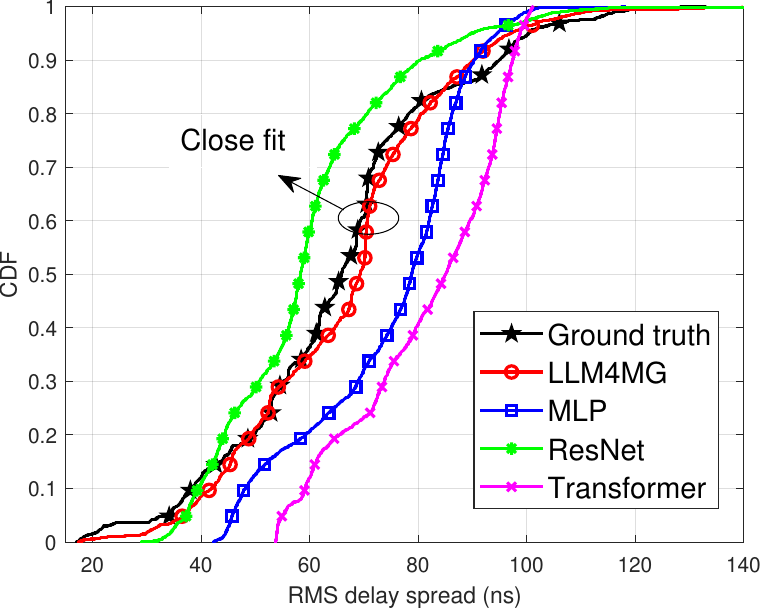}
	\caption{RMS delay spreads via the proposed LLM4MG, MLP, ResNet, and Transformer.}
	\label{RMS_DS}
	\end{figure}

\begin{figure}[!t]
		\centering	\includegraphics[width=0.45\textwidth]{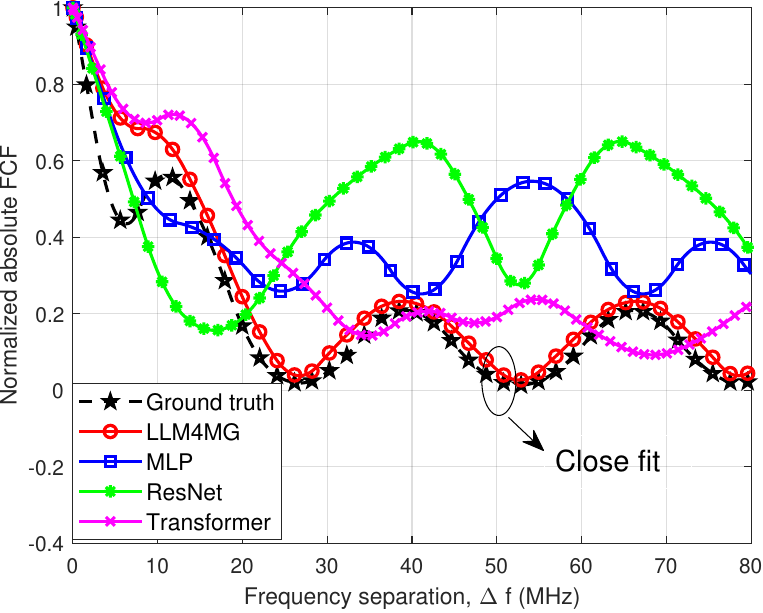}
	\caption{Normalized absolute FCFs via the proposed LLM4MG, MLP, ResNet, and Transformer.}
	\label{FCF}
	\end{figure}

Key channel statistical properties are simulated in Figs.~\ref{PDP}--\ref{FCF} based on the generated multipath power and delay parameters via the proposed LLM4MG and baselines. In Fig.~\ref{PDP}, we obtain time-varying PDPs for spatially-consistent channels using 3 snapshots as an example. A close fit between the RT-based results, i.e., ground truth, and the result based on the proposed LLM4MG is achieved, where time-varying PDPs evolves smoothly  over time in spatially-consistent channels. However, due to the limitations of conventional deep learning models in interference capabilities, the inaccuracies in generated multipath power and delay further lead to significant discrepancies between the generated time-varying PDP and the ground truth. Furthermore, channel spatial consistency cannot be captured via the baselines. In Fig.~\ref{RMS_DS}, the RMS delay spread obtained by the proposed LLM4MG and the ground truth exhibit strong agreement. Nevertheless, the RMS delay spread obtained by the conventional deep learning models is larger than the ground truth. In Fig.~\ref{FCF}, a close fit between FCFs obtained by the proposed LLM4MG and the ground truth is also demonstrated, whereas the conventional deep learning models exhibit a significant deviation from the ground truth.  Therefore, Figs.~\ref{RMS_DS} and \ref{FCF} show that the multipath power and delay generated by the conventional deep learning models incorrectly assess the delay spread characteristics and channel frequency selectivity.

  
\subsubsection{Generalization Experiments}
Generalization, i.e., the capability of the model to sustain performance in novel cases, is essential for real-world deployment, as it minimizes the necessity for the frequent update. To validate the generalization capability of the proposed LLM4MG, we evaluate the model trained on the SynthSoM-V2I dataset under urban scenario with low VTD at mmWave by testing its performance via few-shot fine-tuning across three additional cases, i.e., urban scenario with high VTD at mmWave, urban scenario with low VTD at sub-6 GHz, and suburban scenario with low VTD at mmWave. Therefore, the cross-VTD, cross-band, and cross-scenario generalization evaluations are conducted. In the generalization experiment, we evaluate the generated multipath power and delay under NMSE. On one hand, the sensitivity of NMSE to outliers more effectively reflects the generalization performance of multipath power/delay generation with small values compared to NMAE. On the other hand, since the multipath generation task is more complex than the LoS/NLoS classification task, the improvement in performance with the proposed LLM4MG is more pronounced compared to conventional deep learning models.

\begin{figure*}[!t]
	\centering
      \subfigure[]{\includegraphics[width=0.49\textwidth]{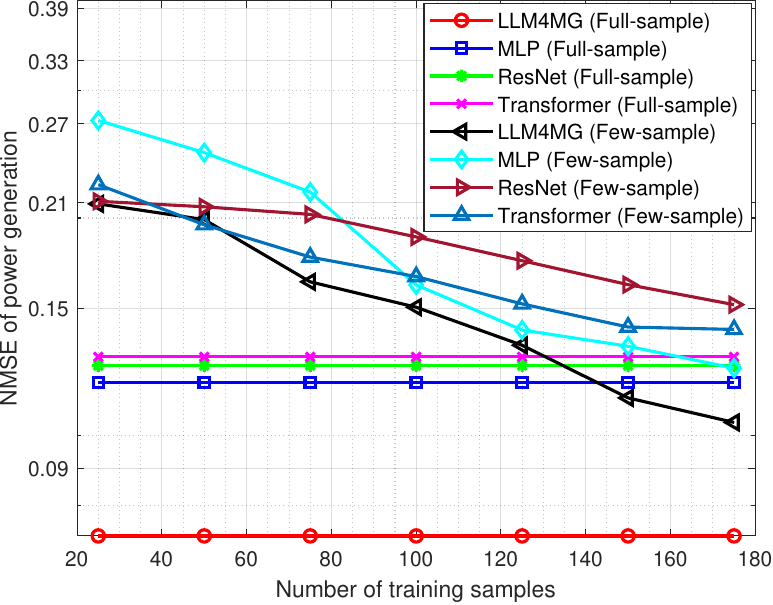}}
 \subfigure[]{\includegraphics[width=0.49\textwidth]{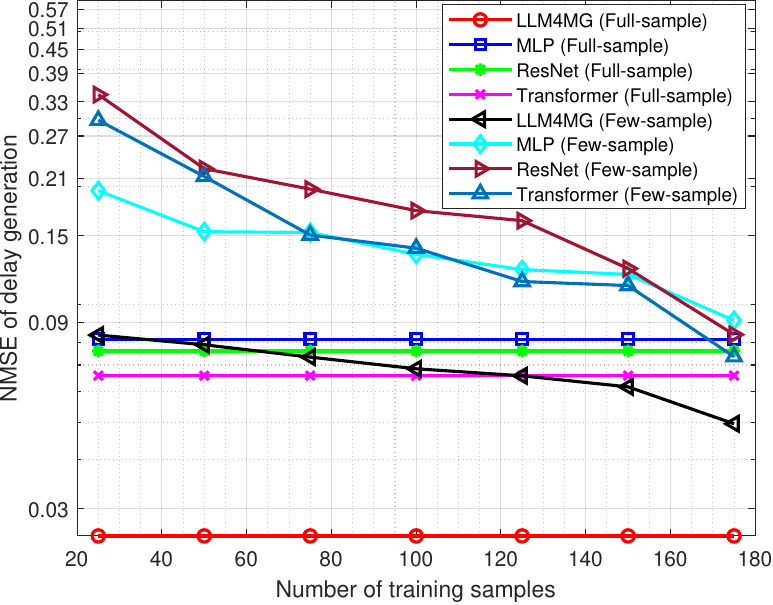}}
   \caption{Generalization performance of the proposed LLM4MG, MLP, ResNet, and Transformer from low VTDs to high VTDs. (a) NMSE of power generation. (b) NMSE of delay generation. }
 \label{to_high}
\end{figure*}
Figs.~\ref{to_high}(a) and (b) illustrate the multipath power generalization performance and multipath delay generalization performance of the proposed LLM4MG, MLP, ResNet, and Transformer from \emph{low VTDs} to \emph{high VTDs}, respectively. Compared to low VTDs, high VTDs contain more dynamic vehicles, resulting in highly time-varying vehicular channels with more pronounced multipath effects. Consequently, cross-VTD generalization testing from low VTDs to high VTDs presents significant challenges. In contrast, the proposed LLM4MG demonstrates superior generalization capability in novel VTDs. The proposed LLM4MG solely needs less than $1.4\%$ of the training samples to achieve the full-shot generation performance of the conventional deep learning models with the optimal performance.

\begin{figure*}[!t]
	\centering
      \subfigure[]{\includegraphics[width=0.49\textwidth]{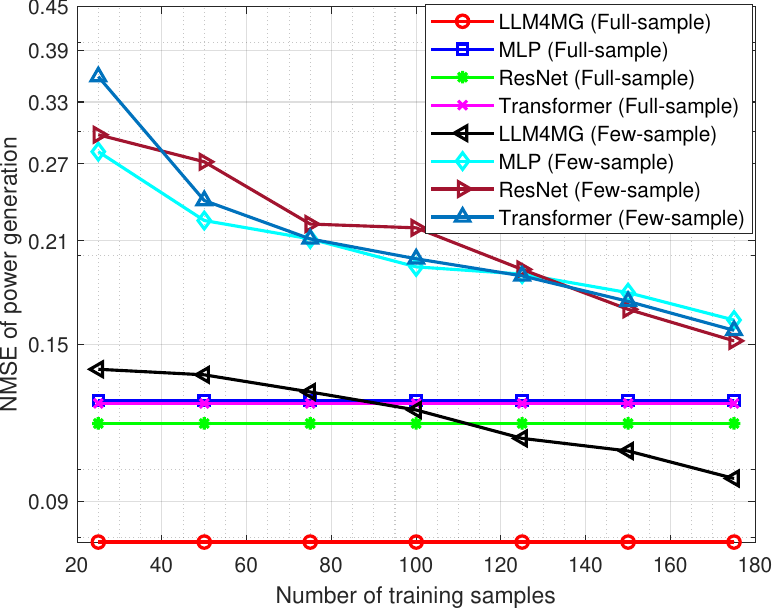}}
 \subfigure[]{\includegraphics[width=0.49\textwidth]{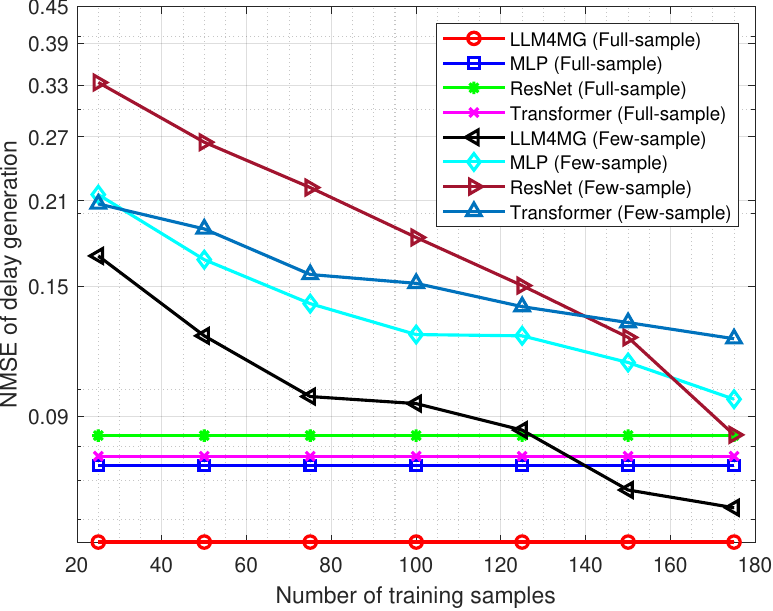}}
   \caption{Generalization performance of the proposed LLM4MG, MLP, ResNet, and Transformer from $60$~GHz to $5.9$~GHz. (a) NMSE of power generation. (b) NMSE of delay generation. }
 \label{to_sub6}
\end{figure*}

The multipath power generalization performance and multipath delay generalization performance of the proposed LLM4MG, MLP, ResNet, and Transformer from \emph{mmWave band ($60$~GHz)} to \emph{sub-6 GHz band ($5.9$~GHz)} are depicted in Figs.~\ref{to_sub6}(a) and (b), respectively. Certainly, variations in carrier frequency bands induce significant changes in multipath power and delay \cite{cai2}, thus rendering cross-band generalization testing from mmWave band to sub-6 GHz band challenging. Similarly, the proposed LLM4MG demonstrates superior generalization performance at the novel frequency band compared to conventional deep learning models. For multipath power generation, the proposed LLM4MG achieves performance comparable to the full-shot optimal conventional deep learning model while requiring fewer than $1.3\%$ of the training samples. The channel with the sub-6 GHz band exhibits richer multipath propagation than that under the mmWave band, resulting in more pronounced delay variations. For multipath delay generation, the NMSE of the proposed LLM4MG for cross-band generalization is larger than that for cross-VTD generalization.
Overall, the proposed LLM4MG achieves performance equivalent to the full-shot optimal conventional deep learning model using $1.4\%$ of the training samples.

\begin{figure*}[!t]
	\centering
      \subfigure[]{\includegraphics[width=0.49\textwidth]{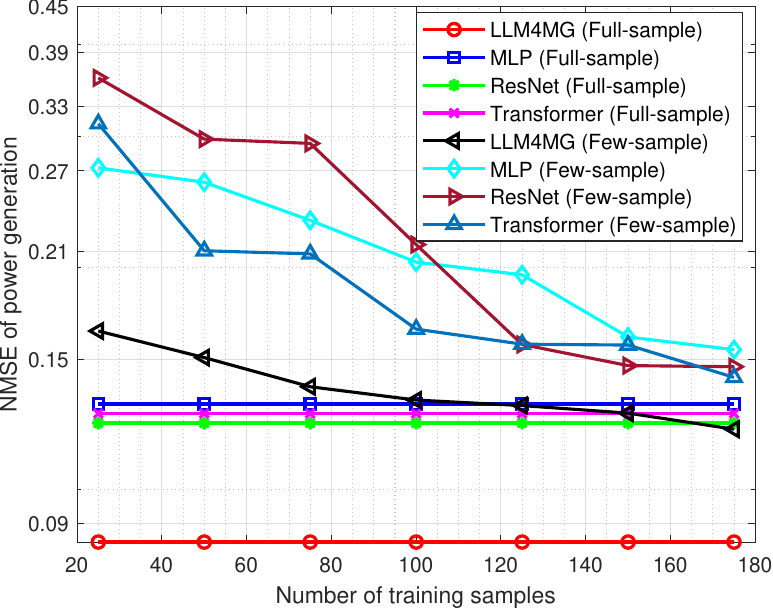}}
 \subfigure[]{\includegraphics[width=0.49\textwidth]{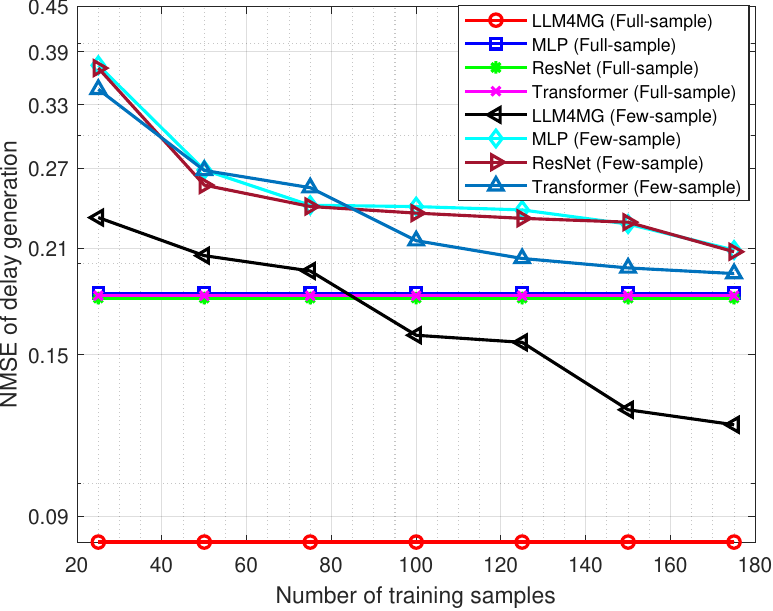}}
   \caption{Generalization performance of the proposed LLM4MG, MLP, ResNet, and Transformer from urban to suburban. (a) NMSE of power generation. (b) NMSE of delay generation. }
 \label{to_suburban}
\end{figure*}

Figs.~\ref{to_suburban}(a) and (b) show the multipath power generalization performance and multipath delay generalization performance of the proposed LLM4MG, MLP, ResNet, and Transformer from \emph{urban scenario} to \emph{suburban scenario}, respectively. Since urban and suburban scenarios have significant differences in the distribution of static buildings and the number of dynamic pedestrians, the different propagation environments in suburban and urban result in challenging cross-scenario generalization testing. In Figs.~\ref{to_suburban}(a) and (b), the proposed LLM4MG demonstrates superior generalization
performance at novel scenarios compared to conventional deep learning models. For multipath power and delay generation, the proposed LLM4MG achieves performance comparable to the full-shot optimal conventional deep learning model while requiring fewer than $1.6\%$ and $1\%$ of the training samples, respectively. 

\subsubsection{Efficiency  and Complexity Evaluation}
To evaluate the practical deployment feasibility of the proposed LLM4MG, we compare its training and inference costs against conventional deep learning models, as summarized in Table~\ref{parameters_table}. All experiments are conducted on identical hardware configurations, utilizing a server equipped with 13th Gen Intel(R) Core(TM) i5-13600KF CPU, NVIDIA GeForce RTX4090 D GPU, and 64GB RAM.
For clarity and comparability, Table~\ref{parameters_table} presents the average task performance metrics aggregated across all methods.
By employing LoRA-based parameter-efficient fine-tuning, the backbone LLM in the proposed LLM4MG reduces its trainable parameters to a fraction of those required by conventional deep learning models, which can demonstrate superior training efficiency and remarkable parameter efficiency. For the proposed LLM4MG, its lightweight backbone LLM further ensures training and inference speeds comparable to conventional deep learning models, maintaining a favorable balance between performance and computational cost.
\begin{table*}[!t]
		\centering
        \renewcommand\arraystretch{1.3}
		\caption{Network Parameters (Trainable Parameters/Total Parameters), Training Cost, And Inference Cost Per Batch.}
		\begin{tabular}{|c|c|c|c|c|}
			\hline
		\makecell[c]{\textbf{Metric}}	 &
	\makecell[c]{\textbf{LLM4MG}} &
	\makecell[c]{\textbf{MLP}} 	 &
	\makecell[c]{\textbf{ResNet}}   &
	\makecell[c]{\textbf{Transformer}}   	\\
			\hline
		Parameters (M)	& $0.16/121.81$ & $29.27/29.27$ & $22.71/22.71$ & $20.48/20.48$\\ 
        \hline
        Training time (ms)	& $9.24$ & $8.53$ & $40.82$ & $48.81$\\ 
        \hline
    Interference time (ms)	& $5.92$ & $0.52$ & $4.53$ & $1.84$\\ 
        \hline
		\end{tabular}	
		\label{parameters_table}
	\end{table*}

\subsubsection{Ablation Experiments} 
To evaluate the effectiveness of the proposed module, we conduct the ablation experiment by revising or deleting the existing module in the proposed LLM4MG. For w/o multi-modal testing, we utilize the uni-modal sensory data collected via camera, LiDAR, or mmWave radar. For w/o propagation embedding testing, typical communication characteristics, including carrier frequency, bandwidth, transceiver distance, azimuth and elevation angles of transceiver antennas, are not embedded into LLaMA 3.2. For the backbone LLM module, the variation contains w/o LLM, i.e., removing LLaMA 3.2,  frozen LLM, i.e., freezing pre-trained weights, and w/o pre-train, i.e., randomly initializing pre-trained weights. The result of ablation experiments is listed in Table~\ref{ablation}. In Table~\ref{ablation}, ablation configurations result in performance degradation, thus demonstrating the necessity of utilizing multi-modal sensory data, propagation embedding, and backbone LLM module. Note that, removing the backbone LLM module causes a most substantial performance drop, which verifies its pivotal role in effectively learning/exploring the mapping mechanism between multi-modal sensing and communications for multipath generation.

\begin{table*}[!t]
		\centering
        \renewcommand\arraystretch{1.3}
		\caption{Testing Results of Ablation Experiments on the Utilization of Multi-Modal Sensing, Propagation Embedding, and Backbone LLM Modules, Where \textbf{Boldface} Indicates The Best Result And \underline{Underlining} Denotes The Second-Best Result.}
        \begin{scriptsize}
		\begin{tabular}{|c|c|c|c|c|c|c|c|c|}
			\hline
		\multirow{2}{*}{\makecell[c]{\textbf{Result}}}	 &
	\multirow{2}{*}{\makecell[c]{\textbf{LLM4MG}}} &
	\multicolumn{3}{c|}{\textbf{w/o Multi-Modal}} 	 &
	\multirow{2}{*}{\makecell[c]{\textbf{w/o Propagation Embedding}}}   &
	\multirow{2}{*}{\makecell[c]{\textbf{w/o LLM}}}  & \multirow{2}{*}{\makecell[c]{\textbf{Frozen LLM}}}  & 	\multirow{2}{*}{\makecell[c]{\textbf{w/o  Pre-train}}}  \\
			\cline{3-5} 
& & Camera & LiDAR & Radar  & & & &    \\
\hline
LoS/NLoS classification accuracy	& $\textbf{92.76}\%$ & $88.75\%$ & $83.83\%$ & ${85.76}\%$& ${89.24}\%$ & ${82.35}\%$ & ${86.34}\%$ & $\underline{89.00}\%$ \\ 
\hline
NMAE of multipath power generation	& $\textbf{0.218}$ & $0.295$ & $\underline{0.275}$ & ${0.279}$& ${0.280}$ & ${0.306}$ & $0.278$ & ${0.280}$ \\ 
        \hline
NMSE of multipath power generation	& $\textbf{0.081}$ & $0.104$ & $\underline{0.096}$ & $\underline{0.096}$& ${0.102}$ & ${0.116}$ & $0.097$ & ${0.099}$ \\ 
        \hline
NMAE of multipath delay generation	& $\textbf{0.099}$ & $0.191$ & ${0.174}$ & ${0.187}$& $\underline{0.167}$ & ${0.204}$ & $0.178$ & ${0.192}$ \\ 
        \hline
NMSE of multipath delay generation	& $\textbf{0.032}$ & $0.064$ & ${0.064}$ & ${0.062}$& $\underline{0.052}$ & ${0.076}$ & $0.053$ & ${0.065}$ \\ 
        \hline
		\end{tabular}	
             \end{scriptsize}
		\label{ablation}
	\end{table*}

\subsection{Utility Validation Through Real-World Generalization Testing via Real2Real, Sim2Real, and Mixed2Real}
The urban scenario with low VTD at mmWave in the SynthSoM-V2I dataset and Scenario 32 in the real-world DeepSense 6G dataset \cite{AlkhateebA} form a twin pair. Scenario 32 in the real-world DeepSense 6G dataset \cite{AlkhateebA} is a typical urban scenario, which is located at the College Ave–5th St intersection in downtown Tempe with the tall building as well as the dense tree surrounding crossroads. In addition, there are many dynamic vehicles and pedestrians.
According to Scenario 32, the DeepSense 6G and SynthSoM-V2I datasets possess spatio-temporally consistent RGB images, LiDAR point clouds, mmWave radar point clouds, and received power. Therefore, we can leverage the real-world DeepSense 6G data in Scenario 32 to evaluate the generalization performance of the proposed LLM4MG, which has been trained on the synthetic SynthSoM-V2I dataset. However,  the DeepSense 6G dataset cannot directly validate the utility of the proposed LLM4MG due to two key gaps, i.e., the absence of vehicle-side data and the lack of depth images. To address these challenges, we employ three approaches, including Real2Real, Sim2Real, and Mixed2Real, to validate the utility of the proposed LLM4MG.
\begin{itemize}
    \item \emph{Real2Real}: The input is real-world multi-modal sensory data at the BS side in the DeepSense 6G dataset to train a non-pre-trained model. The NMSE of Real2Real testing between the generated power and the real-world power is calculated. Real2Real testing provides a performance baseline, given that non-pre-trained model is employed.
    \item \emph{Sim2Real}: The input is real-world multi-modal sensory data at the BS side in the DeepSense 6G dataset to train the model, which has been trained on the synthetic SynthSoM-V2I dataset. The NMSE of Sim2Real testing between the generated power and the real-world power is calculated.  Sim2Real testing analyzes the performance gain of utilizing models trained on the synthetic SynthSoM-V2I dataset.
    \item \emph{Mixed2Real}: The input is mixed, i.e., real-world  and synthetic, multi-modal sensory data at the vehicle and BS sides in the DeepSense 6G dataset and the SynthSoM-V2I dataset, where the ratio of real-world data to synthetic data is~$1:1$. Specifically, there  are real-world RGB images/LiDAR point clouds/mmWave radar point clouds at the BS side, synthetic depth images at the BS side, and synthetic RGB images/depth images/LiDAR point clouds/mmWave radar point clouds at the vehicle side. The mixed multi-modal sensory data is utilized to  train the model, which has been trained on the synthetic SynthSoM-V2I dataset. The NMSE of Mixed2Real testing between the generated power and the real-world  power is calculated. Mixed2Real testing capitalizes on the performance gain achieved by combining real-world and synthetic data.
\end{itemize}

\begin{figure}[!t]
		\centering	\includegraphics[width=0.495\textwidth]{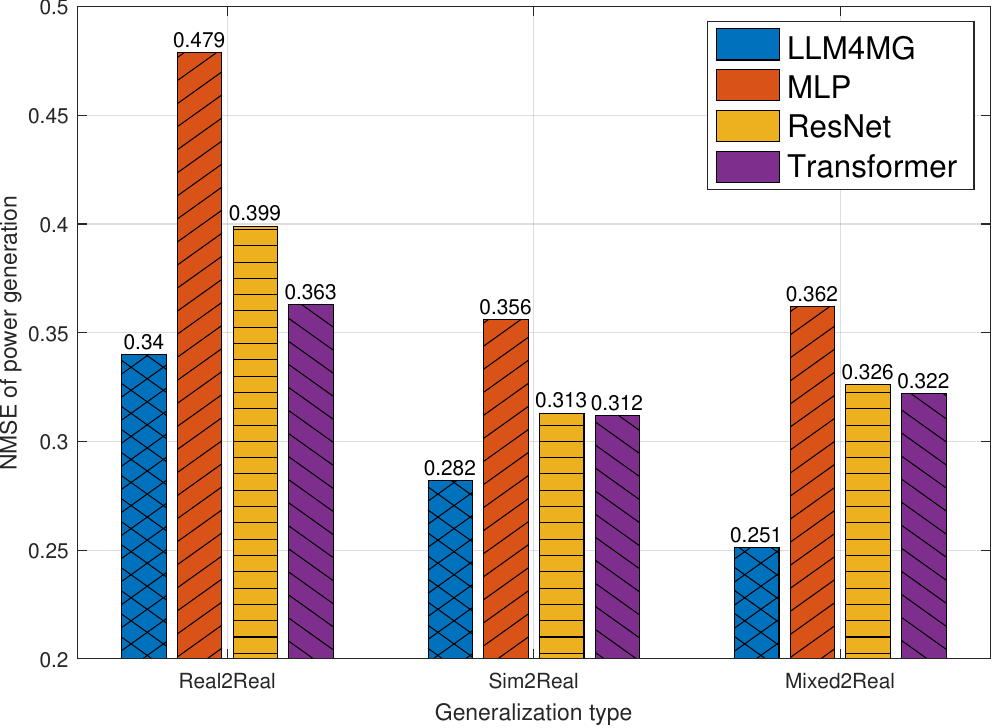}
	\caption{Generalization performance of the proposed LLM4MG, MLP, ResNet, and Transformer via Real2Real, Sim2Real, and Mixed2Real testing.}
	\label{measurement}
	\end{figure}

Fig.~\ref{measurement} depicts the generalization performance of LLM4MG, MLP, ResNet, and Transformer via Real2Real, Sim2Real, and Mixed2Real testing. Across all three validation approaches, the proposed LLM4MG demonstrates significantly superior performance compared to conventional deep learning models. Owing to more significant distributional discrepancy between synthetic and real-world data, the proposed LLM4MG exhibits higher NMSE in real-world generalization testing than in cross-VTD, cross-band, and cross-scenario generalization testing. As anticipated, by exploiting pre-trained LLMs and incorporating synthetic data, the results demonstrate that the  Sim2Real performance  exceeds  Real2Real performance whereas remains lower than Mixed2Real performance.
    
\subsection{High-Precision Multipath Generation Necessity Demonstration Through Channel
Capacity Comparison}
To assess how multipath generated by the proposed LLM4MG versus the conventional deep learning models influence system design and demonstrate the necessity of high-precision multipath generation, we compare the channel capacities computed from their respective generated multipath parameters.
In system design, channel capacity is the theoretical upper bound on the amount of information that can be transmitted over a channel with an arbitrarily low probability of error, assuming optimal coding and modulation. In our evaluation, we consider a system with bandwidth $B$, which is divided into $\mathcal{S}=\{1, \ldots, S\}$ segments/subcarriers within which the channel can be regarded as flat. For the $s$-th bandwidth segment, the SNR $\gamma_{s}$  can be written as 
\begin{equation}
\gamma_{s}=\frac{P_s}{N_0 B S^{-1}}
\label{CC1}
\end{equation}
where $P_s$ and $N_0$ denote the received power on the $s$-th bandwidth segment and the noise power spectral density, respectively. Note that $P_s$ is dependent on $s$ due to multipath fading. Consequently, the overall channel capacity is calculated according to Shannon formula, which is expressed by 
\begin{equation}
    C_\text{capacity}=BS^{-1} \sum_{s \in \mathcal{S}} R_{s} 
    \label{CC2}
\end{equation}
with 
\begin{equation}
R_{s}=\log _2\left(1+\gamma_{s}\right).
\label{CC3}
\end{equation}

In the comparison, the generated multipath parameter in the urban scenario with low VTD at sub-6 GHz in the SynthSoM-V2I dataset is considered. The carrier frequency is $f_c=5.9$~GHz with $B = 20$~MHz  bandwidth. The number of bandwidth segments is set to ${S} = 128$ and a typical value of the noise power spectral density, i.e., $N_0=-174$~dBm/Hz. By further assigning random phase within $[0,2\pi)$, we obtain the channel frequency response (CFR) via the Fourier transform of delay. Based on the CFR, the received power on the $s$-th bandwidth segment, i.e., $P_s$, is computed. As as consequence, the overall channel capacity can be calculated as \eqref{CC1}--\eqref{CC3}. 

\begin{figure}[!t]
		\centering	\includegraphics[width=0.495\textwidth]{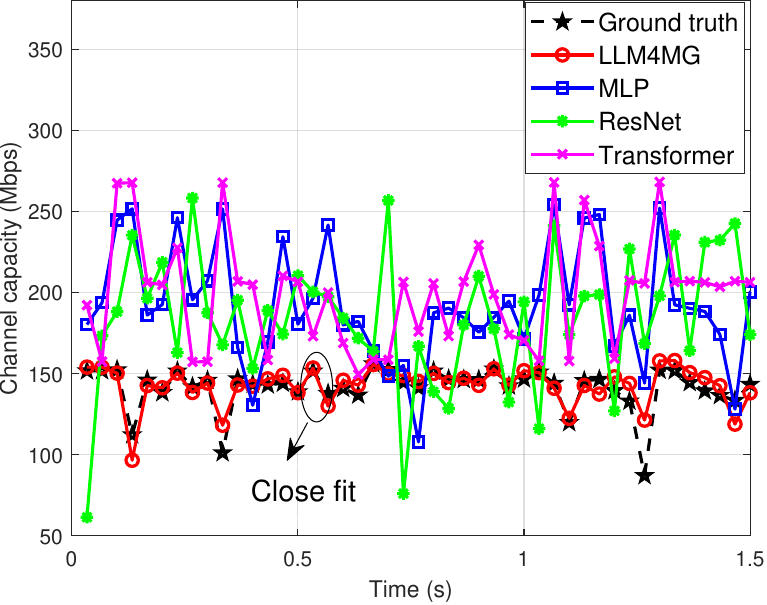}
	\caption{Channel capacities of the proposed  LLM4MG, MLP, ResNet, and Transformer.}
	\label{Capacity}
	\end{figure}

Fig.~\ref{Capacity} compares channel capacities of the proposed LLM4MG, MLP, ResNet, and Transformer. In Fig.~\ref{Capacity}, channel capacities calculated by the RT data and by the generated multipath parameter via the proposed LLM4MG are closely aligned. This also validates the effectiveness of randomly generating multipath phase
via uniform distribution over $[0,2\pi)$.  
By contrast, conventional deep learning models either underestimate or overestimate channel capacities. Using RT data as the ground truth, the channel capacity calculated by the proposed LLM4MG attains an accuracy of 96.20\%, which is more than 30\% higher than the accuracy of  conventional deep learning models. Consequently, the high-precision generation of multipath parameters is of paramount necessity  to the successful  system design.

\section{Conclusions}
Based on a new constructed multi-modal sensing-communication dataset, i.e., SynthSoM-V2I, this paper has proposed a novel LLM4MG. The proposed LLM4MG has adapted LLM for multipath generation via SoM for the first time. To enable the cross-modal generation of fine-grained channel multipath from sensory data, the multi-modal sensing-communication mapping mechanism has been explored by LLaMA 3.2, incorporating LoRA-based parameter-efficient fine-tuning and propagation-aware prompt engineering. Simulation results have shown that the proposed LLM4MG has achieved the best  LoS/NLoS classification accuracy of $92.76\%$. The NMSEs of power and delay generation of the proposed LLM4MG have been $0.099$ and $0.032$, which have achieved over $2.02$ dB and $4.49$ dB enhancement compared to conventional deep learning models, respectively. In the cross-VTD, cross-band, and cross-scenario generalization testing, the proposed LLM4MG has achieved
performance equivalent to full-shot conventional deep learning models using less than $1.6\%$ of the training samples. Furthermore, the utility of the proposed LLM4MG has been verified by real-world generalization evaluation via Real2Real, Sim2Real, and Mixed2Real testing. For the system design, the necessity of high-precision multipath generation has been shown, where channel capacity calculated by the proposed LLM4MG has attained an accuracy of 96.20\% and has been more than 30\% higher than  that of the conventional deep learning model.

\ifCLASSOPTIONcaptionsoff
  \newpage
\fi

\end{document}